\documentstyle[aps,prl,preprint]{revtex}
\tighten
\input epsf
\begin{document}
\draft \title{\bf Metastable states as a key to the dynamics of supercooled liquids }
\author{Stephan B\"uchner and Andreas Heuer}
\address{Max-Planck-Institut f\"ur Polymerforschung, Ackermannweg 10, D-55128 Mainz,
Germany}
\date{\today} \maketitle

\begin{abstract}
Computer simulations of a model glass-forming system are presented, which are
particularly sensitive to the correlation between the dynamics and the
topography of the potential energy landscape. This analysis clearly reveals that
in the supercooled regime the dynamics is strongly influenced by the presence of
deep valleys in the energy landscape, corresponding to long-lived metastable
amorphous states. We explicitly relate non-exponential relaxation effects and
dynamic heterogeneities to these metastable states and thus to the specific
topography of the energy landscape.
\end{abstract}
\ \\
It has been proposed a long time ago that a deeper understanding of relaxation
processes in complex systems can be achieved by analysing the properties of the
potential energy landscape in the high-dimensional configuration space. At
sufficiently low temperatures the properties of the system are mainly dominated
by the local energy minima ({\it inherent structures}) and the dynamics can be
viewed as hopping processes between adjacent inherent structures
\cite{Goldstein69}. Important information like the accessibility of the ground
state in proteins \cite{Shakhnovich91,Klimov96} and clusters \cite{Ball96} or
the relevance of substates in proteins \cite{Frauenfelder91} could indeed be
gained by this approach. Although this approach has been also discussed for
supercooled liquids \cite{Angell95,Stillinger95}, quantitative information is
rare which might help to elucidate the specific properties like non-Arrhenius
temperature behavior or non-exponential relaxation
\cite{Ediger96,Debenedetti96}. For example, no concrete evidence exists for the
relevance of a few specific states in the dynamics of supercooled liquids in
analogy to the substates of proteins.

For approaching this problem via computer simulations, an appropriate diagnostic
tool is to perform molecular dynamics (MD) simulations and to regularly quench
the system, thus mapping the MD trajectory on a set of inherent structures
\cite{Stillinger83}. Along this line Sastry et al. analysed a binary
Lennard-Jones system upon cooling \cite{Sastry98}. They observed that at nearly
the same temperature $T_r$ the structural ($\alpha$) relaxation becomes
non-exponential and the average energy of the inherent structures starts to
decrease. From the presence of a common onset temperature $T_r $ they concluded
that for $T < T_r$ the dynamics is already influenced by the energy landscape
\cite{Sastry98}. However, only for even lower temperatures around the cricital
temperature of the mode coupling theory \cite{Gotze92} the dynamics can indeed
be viewed as hopping between different inherent structures \cite{Schroder99}.

In this Letter we combine the above method, yielding information about the
inherent structures in configuration space, with a simultaneous determination of
the mobility as a measure for the dynamics in real space. Thus we have a unique
way of correlating the topography of the energy landscape with the dynamics in
real space. This approach is, of course, not restricted to the analysis of
supercooled liquids. Without invoking any model assumptions we obtain
unprecedented information about the origin of the complex dynamics in glass
formers. It provides the underlying mechanisms for several empirical
observations recently presented in literature. Apart from the above-mentioned
results by Sastry et al. we particularly refer to the observation of {\it
dynamic heterogeneities}. By a variety of experiments
\cite{Schmidt-Rohr91,Cicerone95a,Schiener96,Richert97a} and simulations
\cite{Hurley95,Heuer97a,Yamamoto98,Donati98,Doliwa98} it has been recently
shown, that the presence of dynamic heterogeneities, corresponding to a
distribution of mobilities, is one of the key features to understand the origin
of the non-exponentiality of the $\alpha$ relaxation.

In order to optimize the information about the energy landscape the system size
has to be carefully chosen. The experimental observation of finite lengthscales
at the glass transition \cite{Donth96b,Tracht98b,Russell98} indicates that a
system with a very large number $N$ of particles can be decomposed into only
weakly interacting subsystems with little correlation among each other.
Therefore the information content of the total potential energy of large systems
about the topography of the energy landscape is limited because it is a sum of
only weakly correlated contributions. Motivated by this simple consideration we
analysed rather small systems which, however, are large enough that finite size
effects are small.

We performed MD simulations at constant density and temperature for a binary
Lennard-Jones-type system with Lennard-Jones parameters and density as in Ref.
\cite{Sastry98,Kob95} but together with a short-range cutoff
\cite{Stillinger83}, thus shifting the temperature scale by ca. 30\% to higher
temperatures as compared to \cite{Sastry98}. Standard Lennard-Jones units and
integration time steps are used. Results are presented for two representative
temperatures $T_l=0.667$ and $T_h=1.667$, which have been chosen such that only
for the lower temperature the typical dynamic features of supercooled liquids,
i.e. the fast decay to a plateau (fast $\beta$-relaxation) and the final
non-exponential decay ($\alpha$-relaxation) of the incoherent scattering
function $S(q_{m},t)$ do appear. $S(q_{m},t)$ is defined as the average of
$S(q_{m},t_0,t)\equiv (1/N)\sum_i
\cos [ \vec{q}_{m}(\vec{r}_i(t+t_0) -\vec{r}_i(t_0))]$ over all
times $t_0$  ($\vec{r}_i(t)$:  position of particle $i$ at time $t$). It
reflects the dynamics of the system on length scale $\pi/q_{m}$ where
$2\pi/q_{m}$ is the average nearest neighbor distance. The $\alpha$-relaxation
time $t_\alpha$ can be defined via the standard condition $S(q_{m},t_{\alpha})=
1/e$. We checked that in the considered temperature range only for $N < 60$
finite size effects are significant for $S(q_m,t)$. Here we present data for
$N=120$; see Fig.1a.

During equilibrium runs of length $500 t_\alpha$ for $T=T_l$ and $T=T_h$,
respectively, we monitored the inherent structures. Representative parts of the
energies of the inherent structures, denoted $\epsilon(t)$, for both
temperatures are shown in Fig.1b. For $T=T_l$ the data show significant
correlations, disappearing for $T=T_h$. Hence pure inspection of the data
already indicates that the dynamics on the energy landscape dramatically changes
upon decreasing the temperature. Interestingly, it turns out that some inherent
structures are revisited several times during some periods of time.
Qualitatively, this can be understood if one assumes that some adjacent inherent
structures form a kind of {\it valley} on the energy landscape in which the
system is caught for some time; see inset of Fig.2a. Hence one has direct
evidence for the presence of long-lived metastable states. For quantification of
the presence of valleys we determined the time intervals $\{[t_i,t^\prime_i]\}$,
during which the system is in a single valley and thus switches between a finite
set of inherent structures until finally it leaves this part of the
configuration space. Formally these time intervals can be defined by the
following conditions: (i) each $[t_i,t^\prime_i]$ is the time
 interval of maximum length such that for every $t_s \in               [t_i,t^\prime_i[$ one has times $t_0
                \in [t_i,t_s]$ and $t_1 \in ]t_s,t^\prime_i]$ with $\epsilon(t_0) =
                \epsilon(t_1)$; (ii) a valley contains at least two different inherent
                structures. Since we want to check the effect of the energy
                landscape on the $\alpha$-relaxation we additionally required  that
                the length of these time intervals is at least $t_{\alpha}/10$.
After determination of the $\{[t_i,t^\prime_i]\}$ we defined effective energies
$\tilde{\epsilon}(t)$ such that during $[t_i,t^\prime_i]$ one identifies
$\tilde{\epsilon}(t)$ with the lowest energy encountered in this interval.
During the remaining times the system can move freely in configuration space and
is not restricted to a valley. We denote these parts of the energy landscape as
the {\it crossover region}. For these time intervals we define
$\tilde{\epsilon}(t)$ as the largest energy of the inherent structures, visited
during this time. In Fig.2a $\tilde{\epsilon}(t)$ is shown for the $\epsilon(t)$
data of Fig.1b for $T
= T_l$. One can clearly see, how the system switches between valleys and crossover
regions.  As already seen in Fig.1, for the high temperature $T_h$ these
features do not appear. Since the average energy of the inherent structures
visited at $T_h$ is much larger than those at $T_l$ it can be rationalised that
valleys, which typically correspond to inherent structures with low energy, are
not visited at high temperatures.

Having quantified relevant properties of the energy landscape one may ask for
the consequences for the single particle dynamics as observed, e.g., by
$S(q_{m},t)$. On a qualitative level, residence in a valley should lead to only
little relaxation. To check this expectation we define the averaged single
particle mobility $\mu(t)$ as the mean square displacement on the time scale of
the $\alpha$-relaxation time, i.e. $\mu(t) \equiv (1/N)
\sum_i [\vec{r}_i(t + t_\alpha/2) -
\vec{r}_i(t - t_\alpha/2)]^2$.  In Fig.2b we show the
time-dependence of $\mu(t)$ for the same data, enabling a comparison between
mobility and location on the energy landscape. The expected relation between
location on the energy landscape and mobility can indeed be observed. This
correlation can be quantified by calculating the average mobility $\langle \mu
\rangle$ in dependence of the  energy $\epsilon$ of the inherent structure.
For $T=T_l$ we separately consider the cases when the system is in a valley or
in the crossover region during the total time interval of length $t_\alpha$,
relevant for the determination of the mobility. The mobility is very small
whenever the system is in a valley; see Fig.3. This clearly shows that the
presence of valleys dramatically slows down the single particle dynamics and is
hence the limiting step for full structural relaxation. Interestingly, the
mobility strongly depends on energy in the crossover regions and is maximum for
large energies. This result might be related to the observation in
Ref.\cite{Sastry98} that at lower temperatures the barriers are higher.  For the
high temperature $T=T_h$ no correlation between energy and mobility is observed,
supporting once again the notion that the topography of the energy landscape is
irrelevant for the dynamics at very high temperatures.

Our results also help to clarify the origin of the dynamic heterogeneities,
giving rise to non-exponential relaxation in supercooled liquids; see above. The
strong correlations, found in Fig.3, directly suggest that the presence of
different mobilities can be attributed to the different locations on the energy
landscape. This point can be made more explicit by considering the function
$S(q_m,t_0,t)$, defined above. In Fig.4 one can see four realisations of
$S(q_m,t_0,t)$. In the extreme cases of very slow/very fast relaxation $t_0$ is
at the beginning of a long time interval in a valley/in a crossover region. As
indicated by the arrow for the case of very slow relaxation, the final
relaxation only takes place after the system has left the valley. Fig.4 shows
that the timescale of relaxation strongly depends on the location on the energy
landscape, finally resulting in a strongly non-exponential decay for the
time-averaged relaxation function $S(q_m,t)$. Hence the structure of the energy
landscape is responsible for at least part of the complexity, observed for the
single particle dynamics.  Interestingly, the decay to the initial plateau
value, i.e. the fast $\beta$ relaxation is independent of the location on the
energy landscape and hence is related to processes for which the global
topography of the energy landscape is not relevant.

Based on the above results it is straightforward to explain the observation in
Ref.\cite{Sastry98} that non-exponentiality is observed in the temperature range
for which the average energy of inherent structures is decreasing with
decreasing temperature. On the one hand, valleys only exist for low-energy
inherent structures, such that the presence of valleys is only relevant at
sufficiently low temperatures; see Fig.2a. On the other hand, the
non-exponentiality is directly connected to the different dynamics in valleys
and crossover regions, respectively; see Fig.4. Combination of both observations
explains why the degree of non-exponentiality and the average energy of inherent
structures are correlated.

As mentioned before, it is crucial to perform this type of analysis for rather
small systems. Based on the definition of valleys, introduced in this work, this
statement can be made more quantitative. Let $p_N$ denote the probability that a
system of size $N$ is in a valley. In the macroscopic limit it is possible to
view a system of size $2N$ as a superposition of two {\it independent} systems
of size $N$. Then the total system is in a valley as long as both subsystems are
in a valley, yielding $p_{2N} = p_N^2$ and thus $p_N = \exp(-N/N_0)$. Thus for
systems with size of less than $N_0$ the properties of the energy landscape and,
correspondingly, the presence of dynamic heterogeneities can be probed by the
present methods. In contrast, for $N \gg N_0$ one has to analyse subvolumes in
order to detect the fluctuations of the inherent structure energy and the
mobility. We mention in passing that alternatively the local fluctuations in
mobility, i.e. the dynamic heterogeneities, can be also monitored by techniques
like analysis of three- or four-time correlation functions \cite{Heuer95}. In
any event, the value of $N_0$ is a measure for the correlation volume,
corresponding to local metastable states. From the present data (as well as from
unpublished data of different system sizes) one can estimate $N_0
\approx 100$ for $T = 0.667$.

In summary, our simulations indicate the relevance of long-lived metastable
structures for the complex dynamics of supercooled liquids, which conceptually
resembles the presence of substates in proteins, separated by high barriers.

We gratefully acknowledge helpful discussions with K. Binder, B. Doliwa, B. D\"unweg, W.
Kob, V. Rostiashvili, R. Schilling, H.W. Spiess, and U. Wiesner. This work was
supported by the German Research Foundation.

\setcounter{figure}{0}
\begin{figure}
\epsfxsize=6.0in
\epsfbox{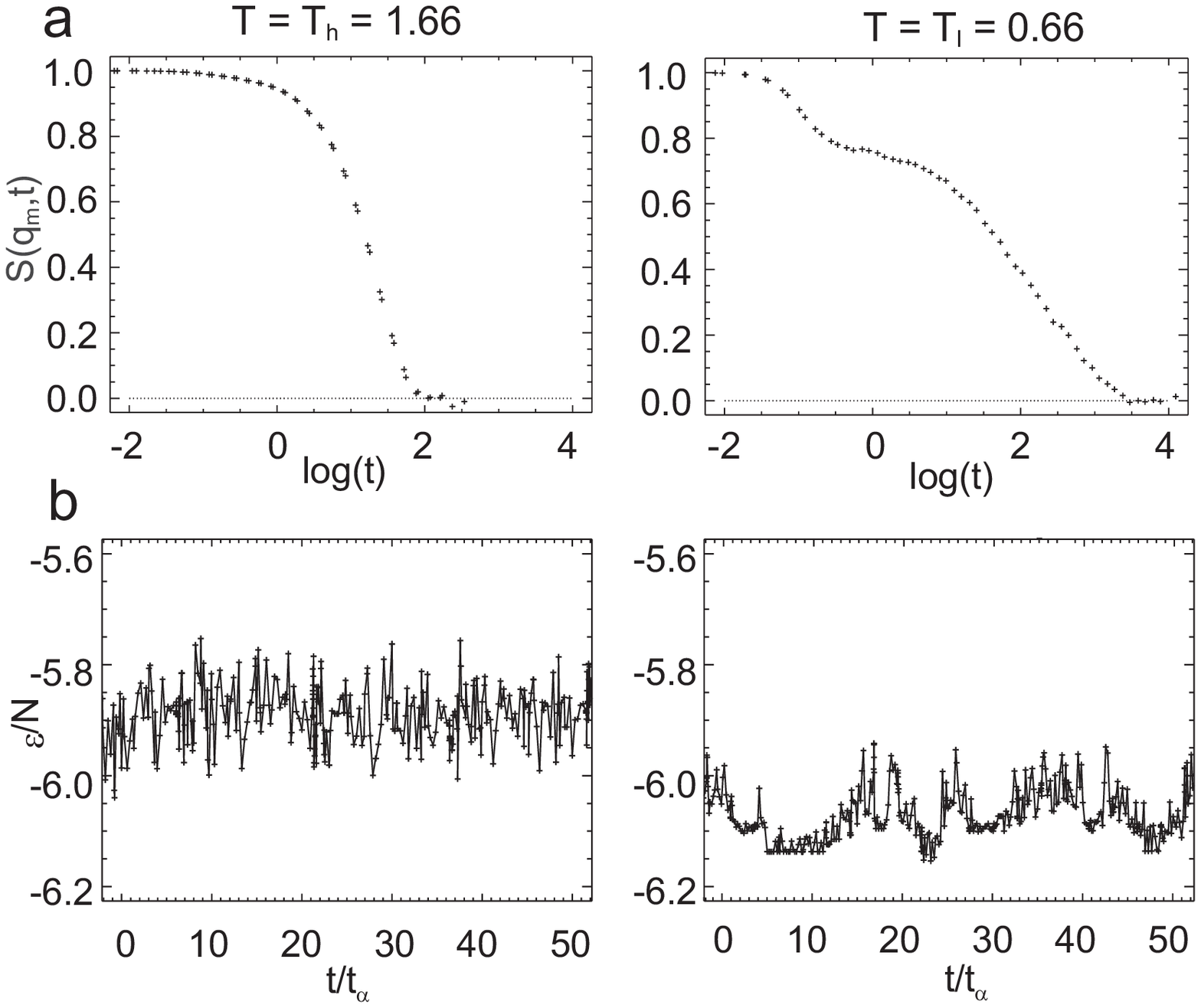}
\vspace{2cm}
\caption{
(a) The incoherent scattering function $S(q_{m},t)$ and (b) the
time series $\epsilon(t)$ for two temperatures $T = T_h
= 1.66$ and $T = T_l = 0.667$. The time axis in (b) has been normalised by
the $\alpha$ relaxation time $t_\alpha$.}
\label{fig:fig1}
\end{figure}
\newpage
\begin{figure}
\epsfxsize=4in
\epsfbox{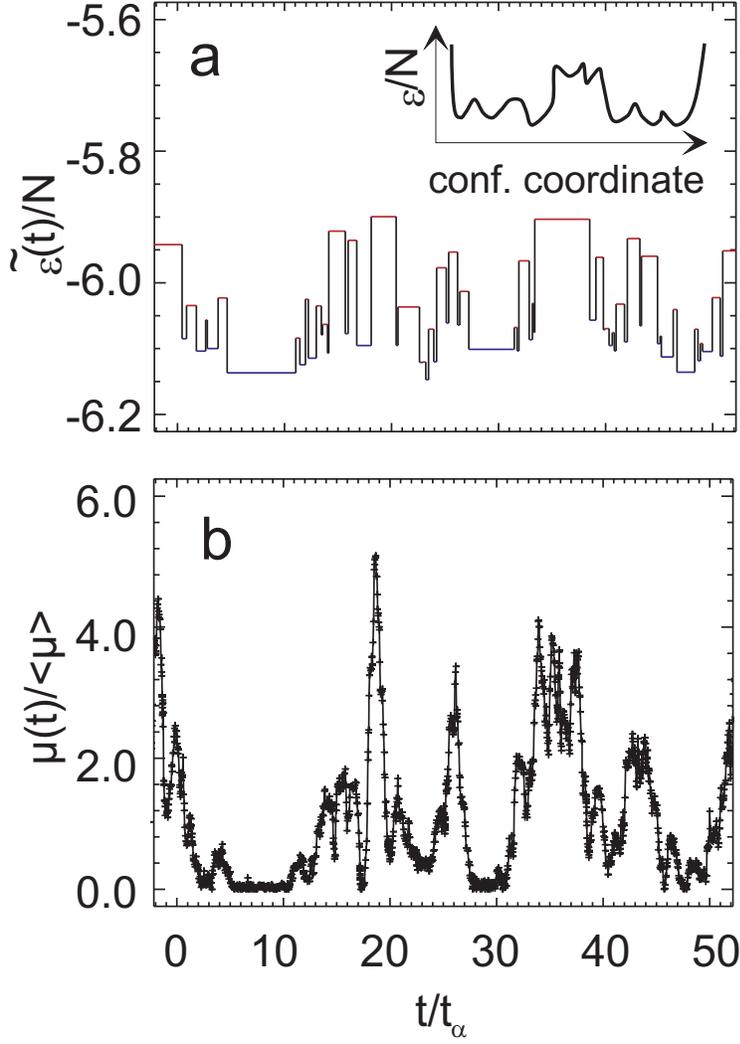}
\vspace{2cm}
\caption{
(a) The effective energies $\tilde{\epsilon}(t)$ and (b) the
mobilites $\mu(t)$ for $T = T_l$. Furthermore a schematic plot of an energy
landscape along a configurational coordinate is shown as an inset, containing
valleys and crossover regions. The presence of valleys on the energy landscape
gives rise to the large time intervals during which the system jumps between a
small number of inherent structures, as seen from $\epsilon(t)$ and highlighted
in the $\tilde{\epsilon}(t)$ representation. During the longest time interval of
this type around $t/t_\alpha
= 8$, the system switches, e.g.,  between 16 different inherent structures.
}
\label{fig:fig2}
\end{figure}
\begin{figure}
\epsfxsize=6.0in
\epsfbox{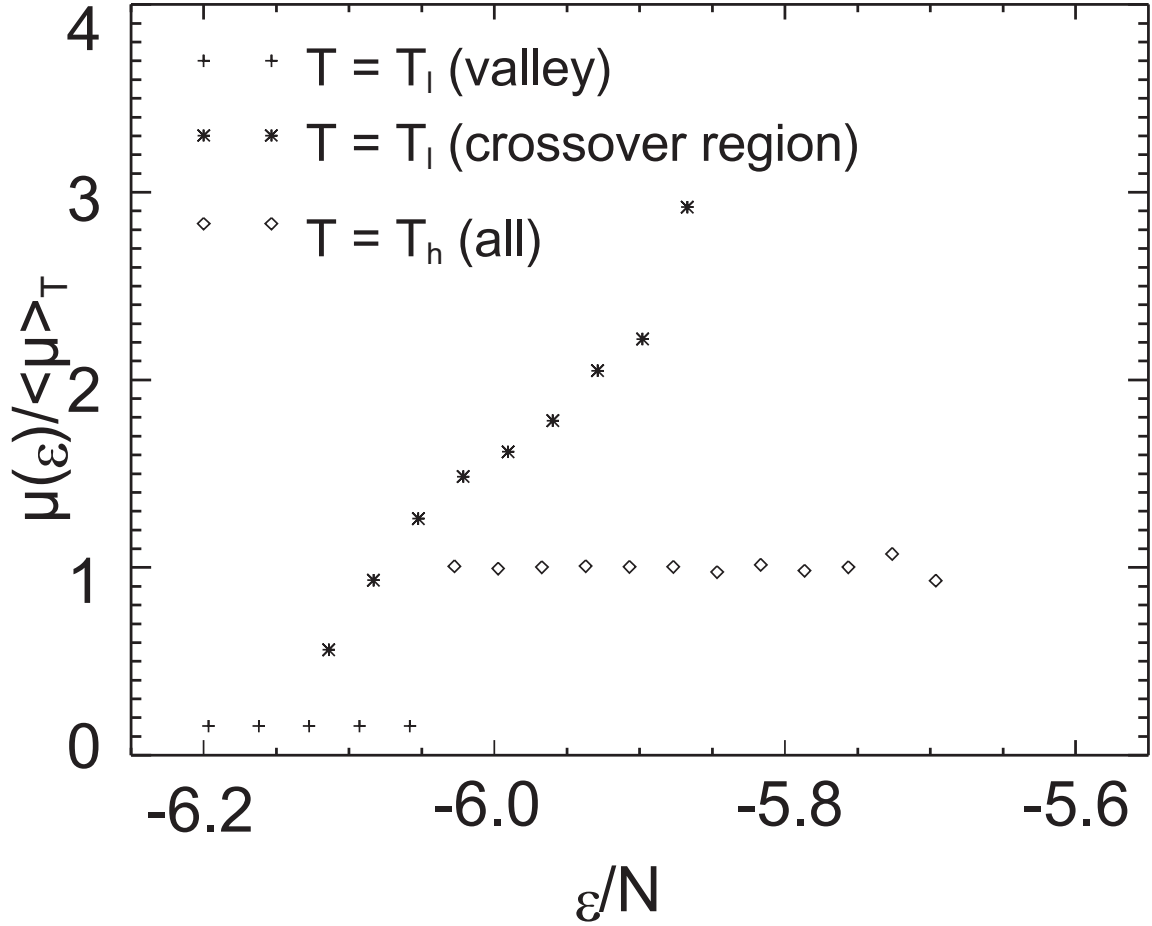}
\vspace{2cm}
\caption{
 The dependence of the average mobility $\mu(\epsilon)$ on the
energy $\epsilon$ of the related inherent structure for $T=T_h$ and $T=T_l$,
individually normalised for both temperatures. For $T=T_l$ we separately
considered the cases that the system was during the total time interval of
length $t_\alpha$, relevant for the definition of the mobility, in a valley or a
crossover region, respectively. }
\label{fig:fig3}
\end{figure}
\begin{figure}
\epsfxsize=6.0in
\epsfbox{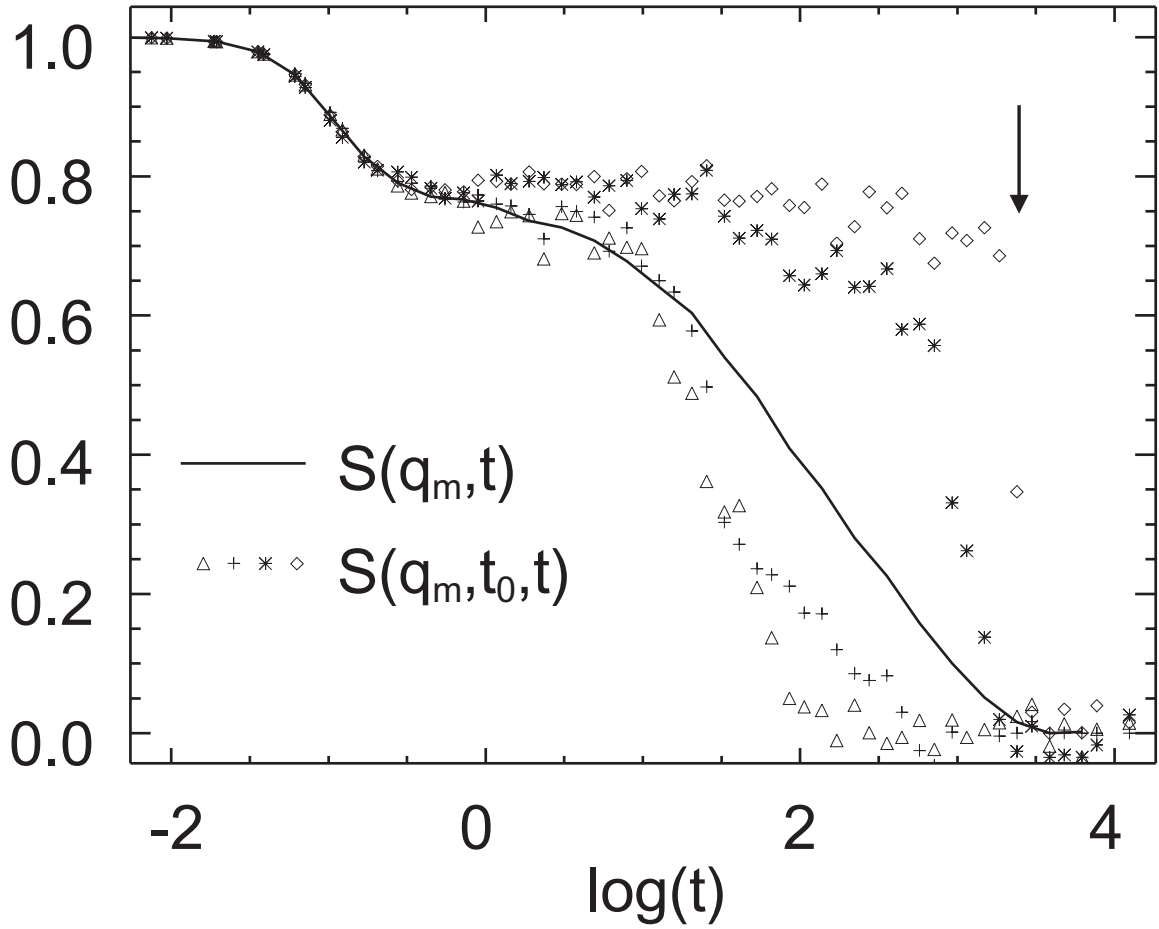}
\vspace{2cm}
\caption{
The incoherent scattering function $S(q_{m},t)$ for $T=T_l$ (same as in Fig.1a)
as well as the function $S(q_{m},t_0,t)$ for different times $t_0$, each
corresponding to a different symbol. For the case of very slow/very fast
relaxation the value of $t_0$ has been chosen such that close to $t_0$ the
system enters for a long time a valley/crossover region. It is obvious that the
non-exponentiality of $S(q_m,t)$ is largely due to a superposition of
$S(q_{m},t_0,t)$ with different decay times. The arrow indicates the time for
the case of the very slow relaxation when the system leaves the initial valley.
}
\label{fig:fig4}
\end{figure}
\end{document}